\documentclass[journal]{IEEEtran}
\usepackage{amsmath}
\usepackage{cite}
\usepackage{graphicx}
\usepackage{booktabs} %table package
\usepackage{lipsum}% http://ctan.org/pkg/lipsum
\usepackage{amssymb}
\usepackage{amsthm}
\usepackage{eqparbox}
\usepackage{multirow}
\usepackage{amsmath}
\usepackage{graphics,graphicx,subfigure}
\usepackage{amsfonts,amssymb,amsmath}

\newtheorem{thm}{\protect\theoremname}

\newtheorem{cor}{\protect\corollaryname}

\newcommand\numberthis{\addtocounter{equation}{1}\tag{\theequation}}
\makeatother

\providecommand{\corollaryname}{Corollary}
\providecommand{\lemmaname}{Lemma}
\providecommand{\propositionname}{Proposition}
\providecommand{\remarkname}{Remark}
\providecommand{\theoremname}{Theorem}

\ifCLASSINFOpdf
\else
\fi
\begin{document}

\begin{LARGE}
This work has been submitted to the IEEE Photonics Technology Letters for possible publication.

Copyright may be transferred without notice, 
after which this version may no longer be accessible.
\end{LARGE}

\newpage
\title{Upper and Lower Bounds for the Ergodic Capacity of MIMO Jacobi Fading Channels}

\author{Amor~Nafkha,~\IEEEmembership{Senior Member,~IEEE,}
        and~R\'emi~Bonnefoi,~\IEEEmembership{Member,~IEEE}% <-this % stops a space
\thanks{A. Nafkha and R. Bonnefoi are with SCEE/IETR research team, CentraleSup\'elec, Avenue de la Boulaie, 35576 Cesson S\'evign\'e, France. E-mail:\{amor.nafkha,Remi.Bonnefoi\}@centralesupelec.fr. A. Nafkha is also with the B-com Institute of Research and Technology, 1219 Avenue des Champs Blancs, 35510 Cesson S\'evign\'e, France}}


\markboth{IEEE Photonics Technology Letters,~Vol.~xx, No.~x, xxxx~2017}%
{Shell \MakeLowercase{\textit{et al.}}}



\maketitle

\begin{abstract}
In multi-(core/mode) optical fiber communication, the transmission channel can be modeled as a complex sub-matrix of the Haar-distributed unitary matrix (complex Jacobi unitary ensemble). In this letter, we present new analytical expressions of the upper and lower bounds for the ergodic capacity of multiple-input multiple-output Jacobi-fading channels. Recent results on the determinant of the Jacobi unitary ensemble are employed to derive a tight lower bound on the ergodic capacity. We use Jensen's inequality to provide an analytical closed-form upper bound to the ergodic capacity at any signal-to-noise ratio (SNR). Closed-form expressions of the ergodic capacity, at low and high SNR regimes, are also derived. Simulation results are presented to validate the accuracy of the derived expressions.
\end{abstract}

\begin{IEEEkeywords}
Ergodic capacity, Multi-(core/mode) optical fiber, space-division multiplexing, Jacobi-fading MIMO channels.
\end{IEEEkeywords}

% \fi
\IEEEpeerreviewmaketitle

\section{Introduction}
\IEEEPARstart{T}o accommodate the exponential growth of data traffic over the last few years, space-division multiplexing (SDM) based on multi-core optical fiber or multi-mode optical fiber \cite{MMF1,MMF2} is expected to overcome the barrier from capacity limit of single-core fiber \cite{SDM}. The main challenge in SDM occurs due to in-band crosstalk between multiple parallel transmission channels (cores or modes). This non-negligible crosstalk can be dealt with by using multiple-input multiple-output (MIMO) signal processing techniques. Assuming important crosstalk between channels (cores or modes), negligible backscattering and near-lossless propagation, we can model the transmission channel as a random complex unitary matrix \cite{Winzer,DFS,Aris}. In \cite{DFS}, authors introduced the Jacobi unitary ensemble to model the propagation channel for fiber-optical MIMO channel and they gave analytical expression for the ergodic capacity. However, to the best of the authors' knowledge, no bounds for the ergodic capacity of the uncorrelated MIMO Jacobi-fading channels exist in the literature so far. The two main contributions of this work are: (\textbf{i}) the derivation of a lower/upper bounds on the ergodic capacity of an uncorrelated MIMO Jacobi-fading channel with identically and independently distributed input symbols, (\textbf{ii}) the derivation of simple asymptotic expressions for ergodic capacity in the low and high SNR regimes.

The rest of this paper is organized as follows: Section \ref{sec.:ProblemFormulation} introduces the MIMO Jacobi-fading channel model and includes the definition of ergodic capacity. We derive a lower and upper bound, at any SNR value, and an approximation, in high and low SNR regimes, to the ergodic capacity in Section \ref{sec.:TightBounds}. The theoretical and the simulation results are discussed in Section \ref{sec.:SimuResults}. Finally, Section \ref{sec.:conclusion} provides the conclusion.

%--------------------------------Problem formulation----------------------------%
\section{Problem formulation}
\label{sec.:ProblemFormulation}
Consider a single segment $m$-channel lossless optical fiber system, the propagation through the fiber may be analyzed through its $2m \times 2m$ scattering matrix given by \cite{Aris}
\begin{equation}
\textbf{S} = 
\begin{bmatrix}
\textbf{R}_{ll} & \textbf{T}_{rl} \\
\textbf{T}_{lr} & \textbf{R}_{rr}
\end{bmatrix}
\label{Scattering_matrix_Aris}
\end{equation}
where $\textbf{T}_{lr}$ and $\textbf{T}_{rl}$ sub-matrices correspond to the transmitted from left to right and from right to left signals, respectively. The $\textbf{R}_{ll}$ and $\textbf{R}_{rr}$ sub-matrices present the reflected signals from left to left and from right to right. Moreover, $\textbf{R}_{ll}=\textbf{R}_{rr}\approx\textbf{0}_{m \times m}$ given the fact that the backscattering in the optical fiber is negligible, and $\textbf{T}=\textbf{T}_{lr}=\textbf{T}_{rl}^\dag$ because the two fiber ends are not distinguishable. The notation $(.)^\dag$ is used to denote the conjugate transpose matrix. Energy conservation principle implies that the scattering matrix $\textbf{S}$ is a unitary matrix ($i.e.$ $\textbf{S}^{-1} = \textbf{S}^\dag$ where the notation $(.)^{-1}$ is used to denote the inverse matrix.). As a consequence, the four Hermitian matrices $\textbf{T}_{lr}\textbf{T}_{lr}^\dag$, $\textbf{T}_{rl}\textbf{T}_{rl}^\dag$, $\textbf{I}_{m}-\textbf{R}_{ll}\textbf{R}_{ll}^\dag$, and $\textbf{I}_{m}-\textbf{R}_{rr}\textbf{R}_{rr}^\dag$ have the same set of eigenvalues $\lambda_1,\lambda_2,....,\lambda_m$. Each of these $m$ transmission eigenvalues is a real number between 0 and 1. Without loss of generality, the transmission matrix $\textbf{T}$ will be modeled as a Haar-distributed unitary random matrix of dimension $m \times m$ \cite{DFS}. 

In this paper, we consider that there are $m_t \leq m$ excited transmitting channels and $m_r \leq m$ receiving channels coherently excited in the input and output side of the $m$-channel lossless optical fiber. Therefore, we only consider a truncated version of the transmission matrix $\textbf{T}$, which we denote by $\textbf{H}$, since not all transmitting or receiving channels may be available to a given link. Without loss of generality, the effective transmission channel matrix $\textbf{H}$ is the $m_r \times m_t$ upper-left corner of the transmission matrix $\textbf{T}$ \cite{Jiang2009}. As a result, the corresponding multiple-input multiple-output channel for this system is given by
\begin{equation}
\textbf{y}=  \textbf{H} \textbf{x} + \textbf{z}
\label{MIMO_input_output_jacobi}
\end{equation}
where $\textbf{y} \in \mathbb{C}^{m_r\times 1}$ is the received signal, $\textbf{x} \in \mathbb{C}^{m_t\times 1}$ is the emitted signal with $\mathbb{E}[\textbf{x}^\dag\textbf{x}]=\frac{\mathcal{P}}{m_t}\textbf{I}_{m_t}$, and $\textbf{z} \sim \mathcal{N}(\textbf{0},\sigma^2\textbf{I}_{m_r})$ is circular-symmetric complex Gaussian noise. We denote $\mathbb{E}[W]$ the mathematical expectation of random variable $W$. The variable $\mathcal{P}$ is the total transmit power across the $m_t$ modes/cores, and $\sigma^2$ is the Gaussian noise variance. We know from \cite{DFS,telatar} that when the receiver has a complete knowledge of the channel matrix, the ergodic capacity is given by

\begin{equation}
	C_{m_t,m_r}^{m,\rho} = \left\{ \begin{array}{rl}
 \mathbb{E}\left[\ln\det\left(\textbf{I}_{m_t} + \frac{\rho}{m_t}  \textbf{H}^\dag\textbf{H}\right)\right] \quad \mbox{ if $m_r\geq m_t$} \\
 \\
 \mathbb{E}\left[\ln\det\left(\textbf{I}_{m_r} + \frac{\rho}{m_t}  \textbf{H}\textbf{H}^\dag\right)\right] \quad \mbox{ if $m_r < m_t$} 
\end{array} \right.
 \label{EC_DFS}
\end{equation}

%\begin{equation}
%C_{m_t,m_r}^{m,\rho} = \mathbb{E}\left[\ln\det\left(\textbf{I}_{m_t} + \frac{\rho}{m_t}  \textbf{H}^\dag\textbf{H}\right)\right]                 
%\label{EC_DFS}
%\end{equation}
where $\ln$ is the natural logarithm function and $\rho=\frac{\mathcal{P}}{\sigma^2}$ is the average signal-to-noise ratio (SNR). 
In this paper, we consider the case where $m_r \geq m_t$ and $m_t+m_r\leq m$. The other case where $m_r < m_t$ and $m_t+m_r\leq m$ can be treated defining $m_t' = m_r$ and $m_r' = m_t$. In the case where $m_t+m_r>m$, it was shown in \cite[Theorem 2,]{DFS} that the ergodic capacity can be deduced from \eqref{EC_DFS} as follows:  
\begin{equation}
C_{m_t,m_r}^{m,\rho} = (m_t+m_r-m) \ln(1+\rho) + C_{m-m_r,m-m_t}^{m,\rho}             
\label{EC_DFS_2}
\end{equation}
The ergodic capacity is defined as the average with respect to the joint distribution of eigenvalues of the covariance channel matrix $\textbf{J}=\frac{1}{m_t}\textbf{H}^\dag\textbf{H}$. The random matrix $\textbf{J}$ follows the Jacobi distribution and its ordered eigenvalues $\lambda_1 \geq \lambda_2 \geq \dots \geq \lambda_{m_t}$ have the joint density given by 
\begin{equation}
\mathcal{F}_{a,b,m}(\lambda)=\chi^{-1}\prod_{1\leq j \leq m_t} \lambda_j^{a} (1-\lambda_j)^{b} V(\lambda)^2
\label{joint_ordred_density_jacobi_fiber}
\end{equation}
where $a=m_r - m_t$, $b=m-m_r-m_t$, $\lambda = (\lambda_1, \dots, \lambda_{m_t})$, $V(\lambda)=\prod_{1 \leq j<k \leq m_t} |\lambda_k-\lambda_j|$, $\chi$ is a normalization constant evaluated using Selberg integral \cite{Kan}, and it is given by:
\begin{equation}
\chi = \prod_{j=1}^{m_t}  \frac{\Gamma(a+1+j)\Gamma(b+1+j)\Gamma(2+j)}{\Gamma(a+b+m_t + j+1)\Gamma(2)}
\label{Evalu_constant_jpdf_jacobi_fiber}
\end{equation}

%---------------------------Tight Bounds----------------------------------------%
\section{Tight bounds on the ergodic capacity}
\label{sec.:TightBounds}
In order to obtain simplified closed-form expressions for the ergodic capacity of the Jacobi MIMO channel, we consider classical inequalities such as Jensen's inequality and Minkowski's inequality. Moreover, we used the concavity property of the $\ln\det(.)$ function given the fact that the channel covariance matrix $\textbf{J}$ is positive definite matrix \cite[Theorem 17.9.1,]{EIT}. 
\vspace{-0.03\linewidth}
\subsection{Upper bound \label{subSec3.1}}
The following theorem presents a new tight upper bound on the ergodic capacity of Jacobi MIMO channel.
\begin{thm}
\label{Tupper}
Let $m_t \leq m_r$, and $m_t + m_r \leq m$, the ergodic capacity of uncorrelated MIMO Jacobi-fading channel, with receiver CSI and no transmitter CSI, is upper bounded by 
\begin{equation}
C_{m_t,m_r}^{m,\rho} \leq m_t \ln\left(1 + \frac{\rho m_r}{m}\right)
\label{Eq1}
\end{equation}
\end{thm}

Proof of Theorem \ref{Tupper}: We propose to use the well known Jensen's inequality \cite{Cvetkovski} to obtain an upper bound for the ergodic capacity. According to this inequality and the concavity of the $\ln\det(.)$ function, we can give a tight upper bound on the ergodic capacity \eqref{EC_DFS} as:
\begin{eqnarray}
C_{m_t,m_r}^{m,\rho} & \leq & m_t \ln\left(1+\rho \mathbb{E} \left[\lambda_1\right]\right)
\label{ergodic_capacity_upperbound}
\end{eqnarray}
Now, the density of $\lambda_1$ is given by \cite[(67),]{DFS} as
\begin{equation}
f_{\lambda_1}(\lambda_1)  = \frac{1}{m_t} \sum_{k=0}^{m_t-1} e_{k,a,b}^{-1} \lambda_1^a (1-\lambda_1)^b \left(P_{k}^{(a,b)}(1-2\lambda_1)\right)^2
\label{density_lambda1}
\end{equation}
where $e_{k,a,b} = \frac{\Gamma(k+a+1)\Gamma(k+b+1)}{k! (2k+a+b+1) \Gamma(k+a+b+1)}$ and $P_{k}^{(a,b)}(x)$ are the Jacobi polynomials \cite[Theorem 4.1.1,]{Ism}. They are orthogonal with respect to the Jacobi weight function $\omega^{a,b}(x) := (1-x)^a (1+x)^b$ over the interval $I=[-1,1]$, where $a,b>-1$, and they are defined by
\begin{equation}
\int^{1}_{-1} (1-x)^a (1+x)^b P_{n}^{(a,b)}(x) P_{m}^{(a,b)}(x) dx = 2^{a+b+1} e_{n,a,b} \delta_{n,m} 
\label{Jacobi_poly_ortho}
\end{equation}
where $\delta_{n,m}$ is the Kronecker delta function. Using \eqref{density_lambda1}, we can write the expectation of $\lambda_1$ as
\begin{equation}
\mathbb{E} \left[\lambda_1\right] =  \sum_{k=0}^{m_t-1} \frac{e_{k,a,b}^{-1}}{m_t} \int^{1}_{0}\lambda_1^{a+1}(1-\lambda_1)^b\left(P_{k}^{(a,b)}(1-2\lambda_1)\right)^2 d\lambda_1
\label{Expectation_1}
\end{equation}
By taking $u= 1-2 \lambda_1$, we can write
\begin{align*}
\mathbb{E} \left[\lambda_1\right] &= \frac{1}{m_t2^{a+b+2}} \sum_{k=0}^{m_t-1} e_{k,a,b}^{-1} \int^{1}_{-1}  (1-u)^{a} (1+u)^b  \\& P_{k}^{(a,b)}(u) \left(P_{k}^{(a,b)}(u)-uP_{k}^{(a,b)}(u)\right) du \numberthis
\label{Expectation_2}
\end{align*}
we recall from \cite[(4.2.9),]{Ism} the following three-term recurrence relation of Jacobi polynomials generation:
\begin{equation}
u P_{k}^{(a,b)}(u) = \frac{P_{k+1}^{(a,b)}(u)}{A_k}  - \frac{C_k P_{k-1}^{(a,b)}(u)}{A_k}  - \frac{B_k P_{k}^{(a,b)}(u)}{A_k}  \: \: \: , k>0
\label{recurrence}
\end{equation}
where $A_k=\frac{(2k+a+b+1)(2k+a+b+2)}{2(k+1)(k+a+b+1)}$, $B_k=\frac{(a^2-b^2)(2k+a+b+1)}{2(k+1)(k+a+b+1)(2k+a+b)}$, and $C_k=\frac{(k+a)(k+b)(2k+a+b+2)}{(k+1)(k+a+b+1)(2k+a+b)}$. Then, by employing \eqref{Jacobi_poly_ortho}, \eqref{Expectation_2}, and \eqref{recurrence}, the expectation of $\lambda_1$ can be expressed as

\begin{align*}
\mathbb{E} \left[\lambda_1\right] &= \sum_{k=0}^{m_t-1} \frac{e_{k,a,b}^{-1}}{m_t2^{a+b+2}} \int^{1}_{-1}  (1-u)^{a} (1+u)^b P_{k}^{(a,b)}(u) \\& \left(P_{k}^{(a,b)}(u)-uP_{k}^{(a,b)}(u)\right) du  \numberthis
\label{Expectation_3}
\end{align*}
thus, we can write
\begin{eqnarray}
\mathbb{E} \left[\lambda_1\right] & = & \frac{1}{2m_t}\sum_{k=0}^{m_t-1} \left(1+\frac{B_k}{A_k}\right) \nonumber \\
                                %  & = &  \frac{a+m_t}{2m_t+a+b}  \nonumber \\
																	& = &  \frac{mr}{m} 
\label{Expectation_4}
\end{eqnarray}
Finally, the upper bound on the ergodic capacity can be expressed as:
\begin{equation}
C_{m_t,m_r}^{m,\rho} \leq m_t \ln\left(1+\frac{\rho m_r}{m}\right)
\label{ergodic_capacity_upperbound_new}
\end{equation}
This completes the proof of Theorem \ref{Tupper}.

In low-SNR regimes, the proposed upper bound expression is very close to the ergodic capacity. Thus, we derive the following corollary.

\begin{cor}
\label{Low_SNR_App}
Let $m_t \leq m_r$, and $m_t + m_r \leq m$. In low-SNR regimes, the ergodic capacity for uncorrelated MIMO Jacobi-fading channel can be approximated as
\begin{equation}
C_{m_t,m_r}^{m,\rho<<<1}  \approx \frac{m_t m_r\rho }{m}
\label{Eq2} 
\end{equation}
\end{cor}

Proof of Corollary \ref{Low_SNR_App}: In low-SNR regimes ($\rho <<< 1$), the function $\ln\left(1+\frac{m_r\rho}{m}\right)$ can be approximated by $\frac{m_r \rho}{m}$.

When the sum of transmit and receive modes, $m_t + m_r$, is larger than the total available modes, $m$, the upper bound expression of the ergodic capacity can be deduced from (\ref{EC_DFS_2}).

\subsection{Lower bound \label{subSec3.2}}

The following theorem gives a tight lower bound on the ergodic capacity of Jacobi MIMO channels.
\begin{thm}
\label{Tlower}
Let $m_t \leq m_r$, and $m_t + m_r \leq m$, the ergodic capacity of uncorrelated MIMO Jacobi-fading channel, with receiver CSI and no transmitter CSI, is lower bounded by 
\begin{equation}
C_{m_t,m_r}^{m,\rho} \geq m_t \ln\left(1+\frac{\rho}{\sqrt[m_t]{F_{m_t,m_r}^{m}}} \right)
\label{Eq3}
\end{equation}
where $F_{m_t,m_r}^{m}=\prod_{j=0}^{m_t-1}\prod_{k=0}^{m-m_r-1} \exp^{\left({\frac{1}{m_r+k-j}}\right)}$

\end{thm}

Proof of Theorem \ref{Tlower}:
We start from Minkowski's inequality \cite{Cvetkovski} that we recall here for simplicity. Let $\textbf{A}$ and $\textbf{B}$ be two $n \times n$ positive definite matrices, then $\det(\textbf{A}+\textbf{B})^{\frac{1}{n}} \geq \det(\textbf{A})^{\frac{1}{n}}+\det(\textbf{B})^{\frac{1}{n}}$
with equality iff $\textbf{A}$ is proportional to $\textbf{B}$. Applying this inequality to \eqref{EC_DFS}, a lower bound of the ergodic capacity can be obtained as
\begin{eqnarray}
C_{m_t,m_r}^{m,\rho} &\geq&  m_t \mathbb{E}\left[\ln\left( 1+\rho \exp^{\left(\frac{1}{m_t}\ln\det\left(\textbf{J}\right)\right)}\right)\right] 
\label{Lower1}
\end{eqnarray}
Recalling that $\ln(1+c \exp^x)$ is convex in $x$ for $x>0$, we apply Jensen's inequality \cite{Cvetkovski} to further lower bound \eqref{Lower1} 
\begin{equation}
C_{m_t,m_r}^{m,\rho} \geq m_t \ln\left( 1+\rho \exp^{\left(\frac{1}{m_t}\mathbb{E}\left[\ln\det\left(\textbf{J}\right)\right] \right)}\right) 
\label{Lower2}
\end{equation}
Using the Kshirsagar's theorem \cite{Kshirsagar}, it has be shown in \cite[Theorem 3.3.3,]{Muirhead}, and \cite{Rouault} that the determinant of the Jacobi ensemble can be decomposed into a product of independent beta distributed variables. We infer from \cite{Rouault} that
\begin{equation}
\ln\det\left(\textbf{J}\right) {\buildrel (d) \over =} \sum_{j=1}^{m_t} \ln T_j
\label{Lower3}
\end{equation}
where ${\buildrel (d) \over =}$ stands for equality in distribution, $T_j$, $j=1,\dots,m_t$ are independent and $
T_j {\buildrel (d) \over =} Beta(m_r-j+1,m-m_r)$
where $Beta(\alpha,\beta)$ is the beta distribution with shape parameters $(\alpha,\beta)$. Taking the expectation over all channel realizations of a random variable $U= \ln\det\left(\textbf{J}\right)$, we get
\begin{equation}
\mathbb{E}\left[U\right] = \sum_{j=0}^{m_t-1} \psi(m_r-j)-\psi(m-j)
\label{Lower5}
\end{equation}
where $\psi(n)$ is the digamma function. For positive integer $n$, the digamma function is also called the Psi function defined as \cite{bookdigamma} 
\begin{equation}
\label{psifunction}
\left\{ \begin{array}{l l}
    \psi(n) = -\gamma  & n=1\\
    \psi(n) = -\gamma + \sum_{k=1}^{n-1} \frac{1}{k}  & n \geq 2
  \end{array} \right.
\end{equation}
where $\gamma\approx 0.5772$ is the Euler-Mascheroni constant. Now, we can finish the proof of the Theorem \ref{Tlower} as follows
\begin{eqnarray}
C_{m_t,m_r}^{m,\rho}  	&\geq& m_t \ln\left( 1+\rho \exp^{\left(\frac{1}{m_t} \sum^{m_t-1}_{j=0} \psi(m_r-j)-\psi(m-j) \right)}\right) \nonumber \\
										%&\geq& \frac{m_t}{\ln(2)} \ln\left( 1+\rho\left(e^{\left(\sum_{j=0}^{m_t-1} \sum_{k=0}^{m-mr-1}\frac{1}{m_r+k-j}\right)}\right)^{\frac{-1}{m_t}}\right) \nonumber \\
										%&\geq& m_t \ln\left(1+\frac{\rho}{\sqrt[m_t]{\prod_{j=0}^{m_t-1}\prod_{k=0}^{m-m_r-1} \exp^{\left({\frac{1}{m_r+k-j}}\right)}}} \right) \nonumber \\
										&\geq& m_t \ln\left(1+\frac{\rho}{\sqrt[m_t]{F_{m_t,m_r}^{m}}} \right)
\label{final}
\end{eqnarray}
where $F_{m_t,m_r}^{m}=\prod_{j=0}^{m_t-1}\prod_{k=0}^{m-m_r-1} \exp^{\left({\frac{1}{m_r+k-j}}\right)}$.
This completes the proof of Theorem \ref{Tlower}.

In high-SNR regimes, the proposed lower bound expression is closed to the ergodic capacity. Thus, we derive the following corollary.

\begin{cor}
\label{High_SNR_App}
Let $m_t \leq m_r$, and $m_t + m_r \leq m$. In high-SNR regimes, the ergodic capacity for uncorrelated MIMO Jacobi-fading channel can be approximated as
\begin{equation}
C_{m_t,m_r}^{m,\rho>>1}  \approx m_t\ln\left( \rho \right)- \sum_{j=0}^{m_t-1} \sum_{k=0}^{m-m_r-1}\frac{1}{m_r+k-j}
\label{Eq4}
\end{equation}
\end{cor}

Proof of Corollary \ref{High_SNR_App}:
In high-SNR regimes ($\rho >> 1$), the function $\ln\left(1+\frac{\rho}{\sqrt[m_t]{F_{m_t,m_r}^{m}}} \right)$ can be approximated by $\ln\left(\rho\right)-\frac{1}{m_t} \ln\left(F_{m_t,m_r}^{m}\right)$.

\section{Simulation results}
\label{sec.:SimuResults}

\begin{figure}[t]
\begin{center}
  \begin{minipage}[b]{0.47\linewidth}
     \includegraphics[width=1.1\linewidth,height=1.0\linewidth]{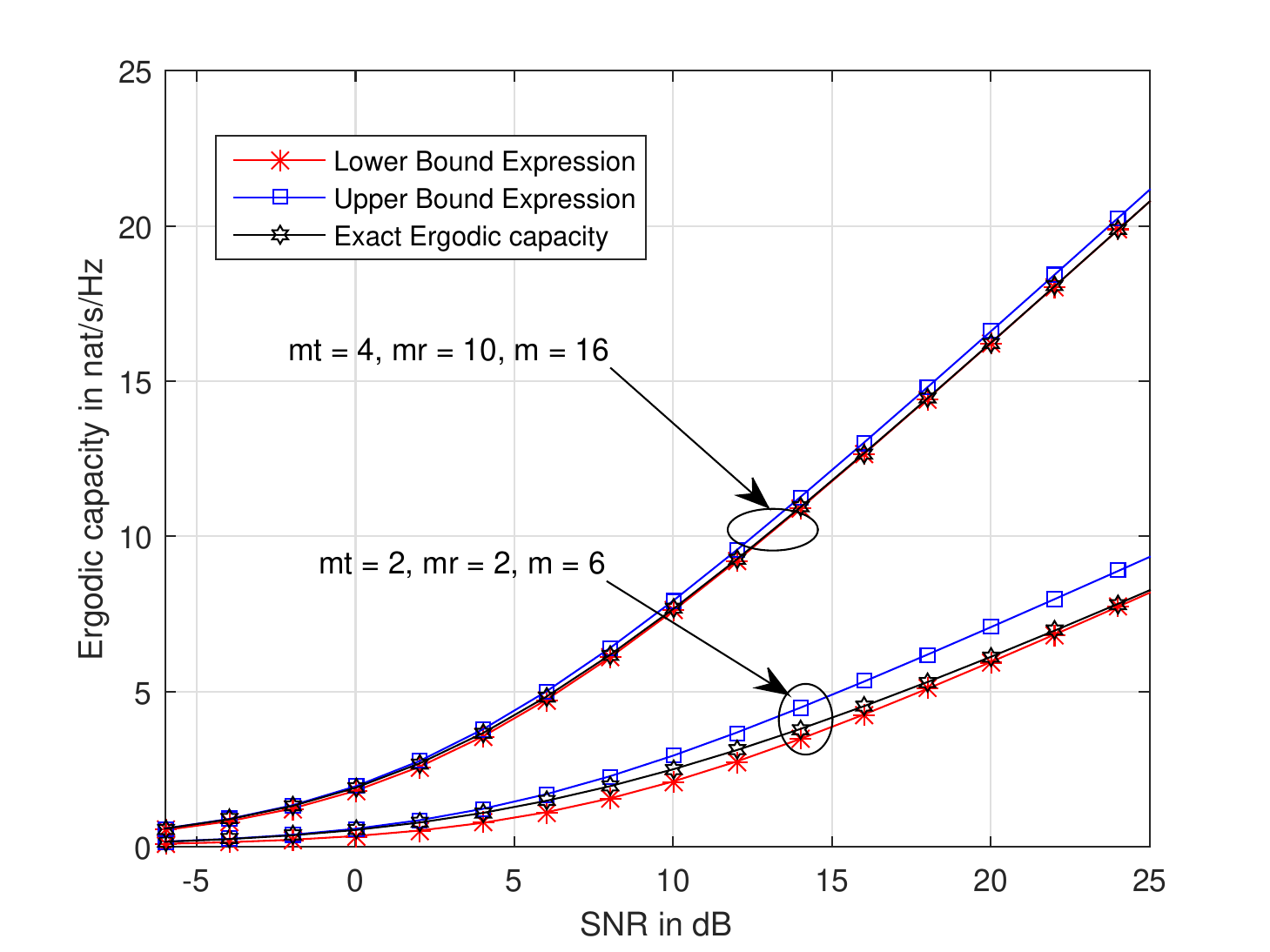}
     %\put(-95,-8){\rotatebox{0}{\mbox{(a)}}}
  \end{minipage}   % no empty line here so the figures are side by side
  \hspace{0.02\linewidth}
  \begin{minipage}[b]{0.47\linewidth}
     \includegraphics[width=1.1\linewidth,height=1.0\linewidth]{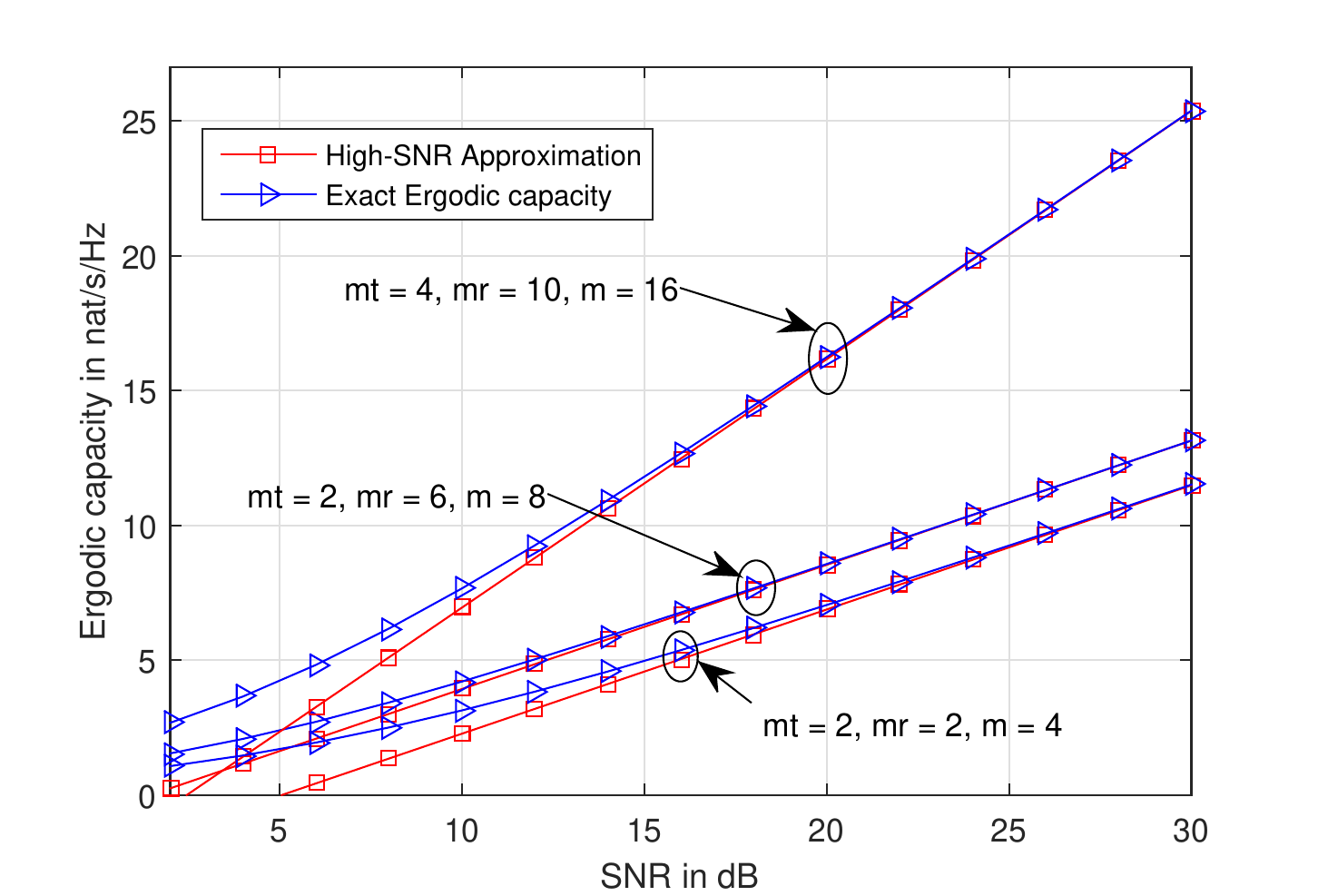}
    % \put(-95,-8){\rotatebox{0}{\mbox{(b)}}}
  \end{minipage}
\end{center}
\caption{{(a)} Comparison of the ergodic capacity and analytical lower-bound and upper-bound expressions for $(m_t=m_r=2, m=6)$, and $(m_t=4, m_r=10, m=16)$ uncorrelated MIMO Jacobi-fading channels, {(b)} High-SNR lower-bound approximation of the ergodic capacity in nats per channel use versus SNR in dB.}
\label{fig:accurate1}
\end{figure}

\begin{figure}[h]
\begin{center}
  \begin{minipage}[b]{0.47\linewidth}
     \includegraphics[width=1.1\linewidth,height=1.0\linewidth]{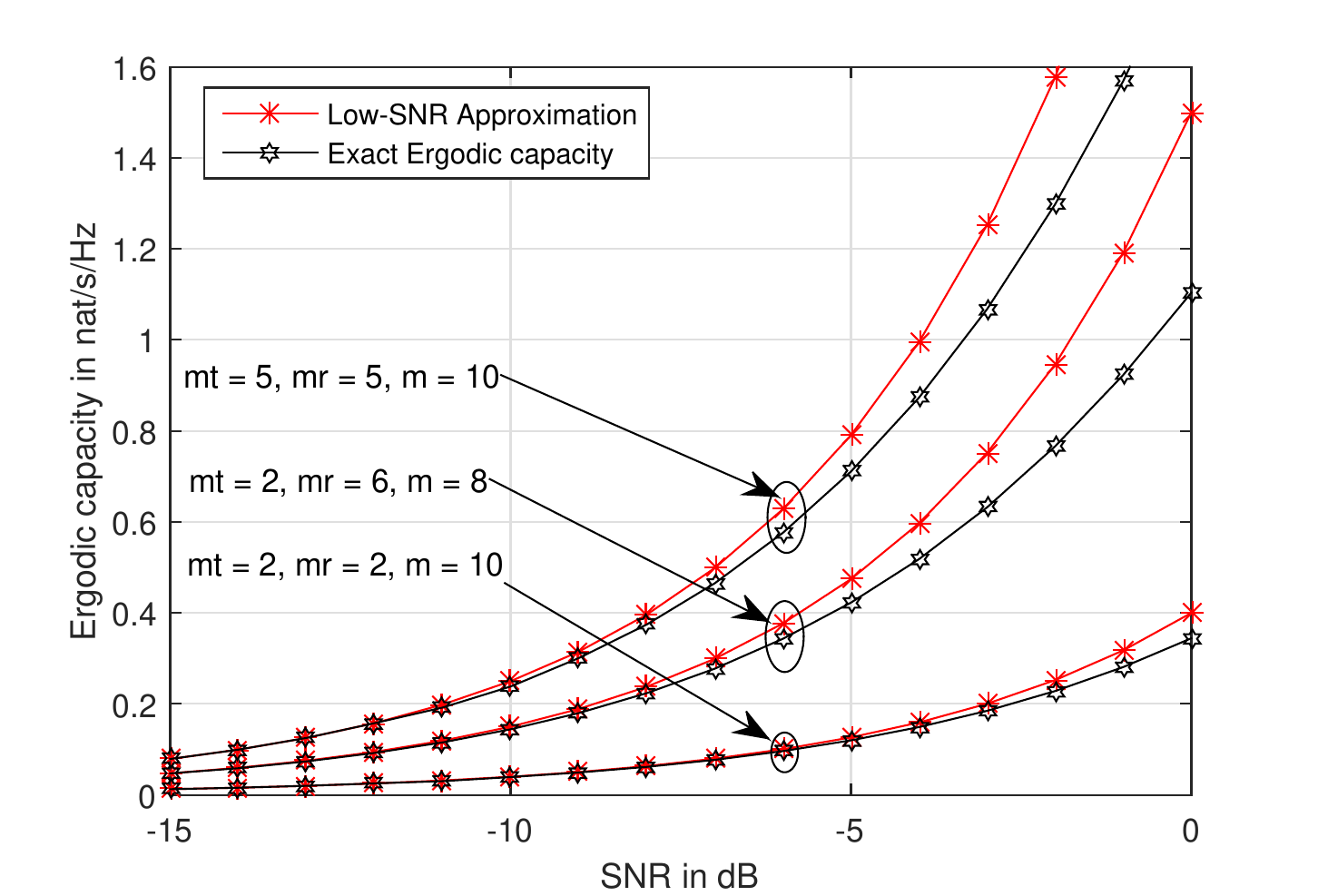}
     %\put(-95,-8){\rotatebox{0}{\mbox{(a)}}}
  \end{minipage}   % no empty line here so the figures are side by side
  \hspace{0.01\linewidth}
  \begin{minipage}[b]{0.47\linewidth}
     \includegraphics[width=1.1\linewidth,height=1.0\linewidth]{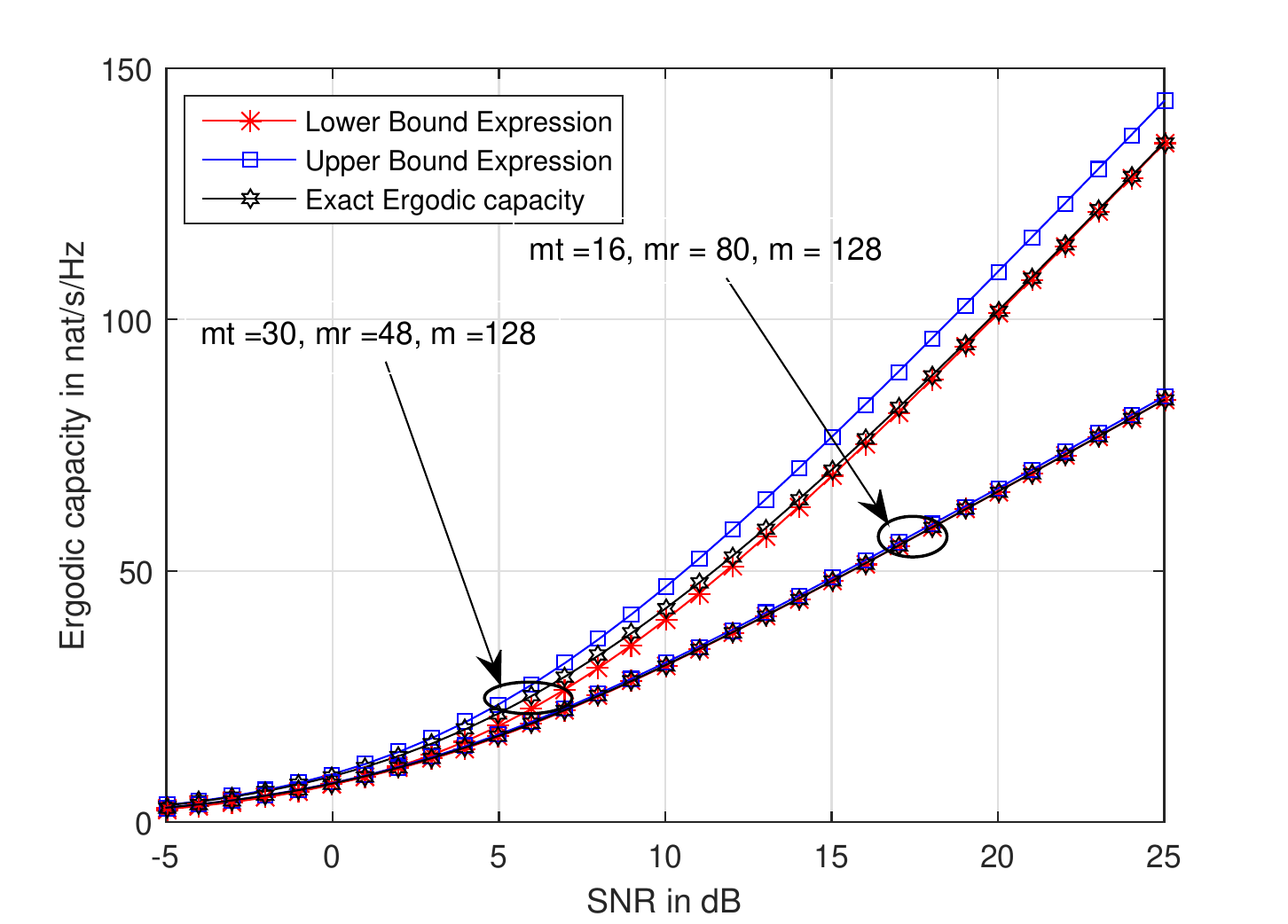}
     %\put(-95,-8){\rotatebox{0}{\mbox{(b)}}}
  \end{minipage}
\end{center}
\caption{{(a)} Low-SNR upper-bound approximation of the ergodic capacity in nats per channel use versus SNR in dB, {(b)} Bounds and simulation results for ergodic capacity of MIMO Jacobi channel capacity, with number available modes $m=128$, for different numbers of transmitting and receiving channels.}
\label{fig:accurate2}
\end{figure}

\begin{figure}[t]
\begin{center}
  \begin{minipage}[b]{0.47\linewidth}
     \includegraphics[width=1.1\linewidth,height=1.0\linewidth]{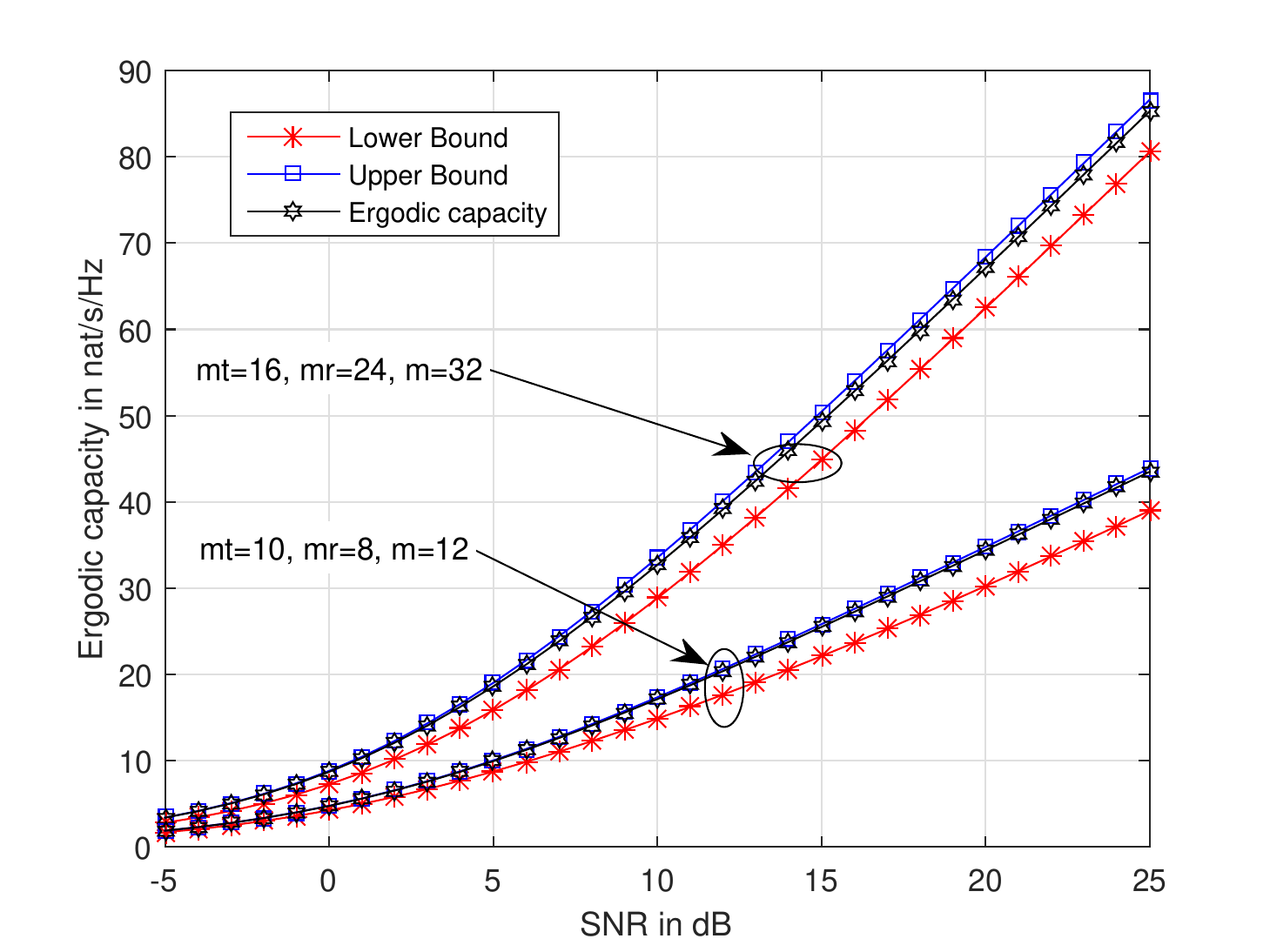}
     %\put(-95,-8){\rotatebox{0}{\mbox{(a)}}}
  \end{minipage}   % no empty line here so the figures are side by side
  \hspace{0.02\linewidth}
  \begin{minipage}[b]{0.47\linewidth}
     \includegraphics[width=1.1\linewidth,height=1.0\linewidth]{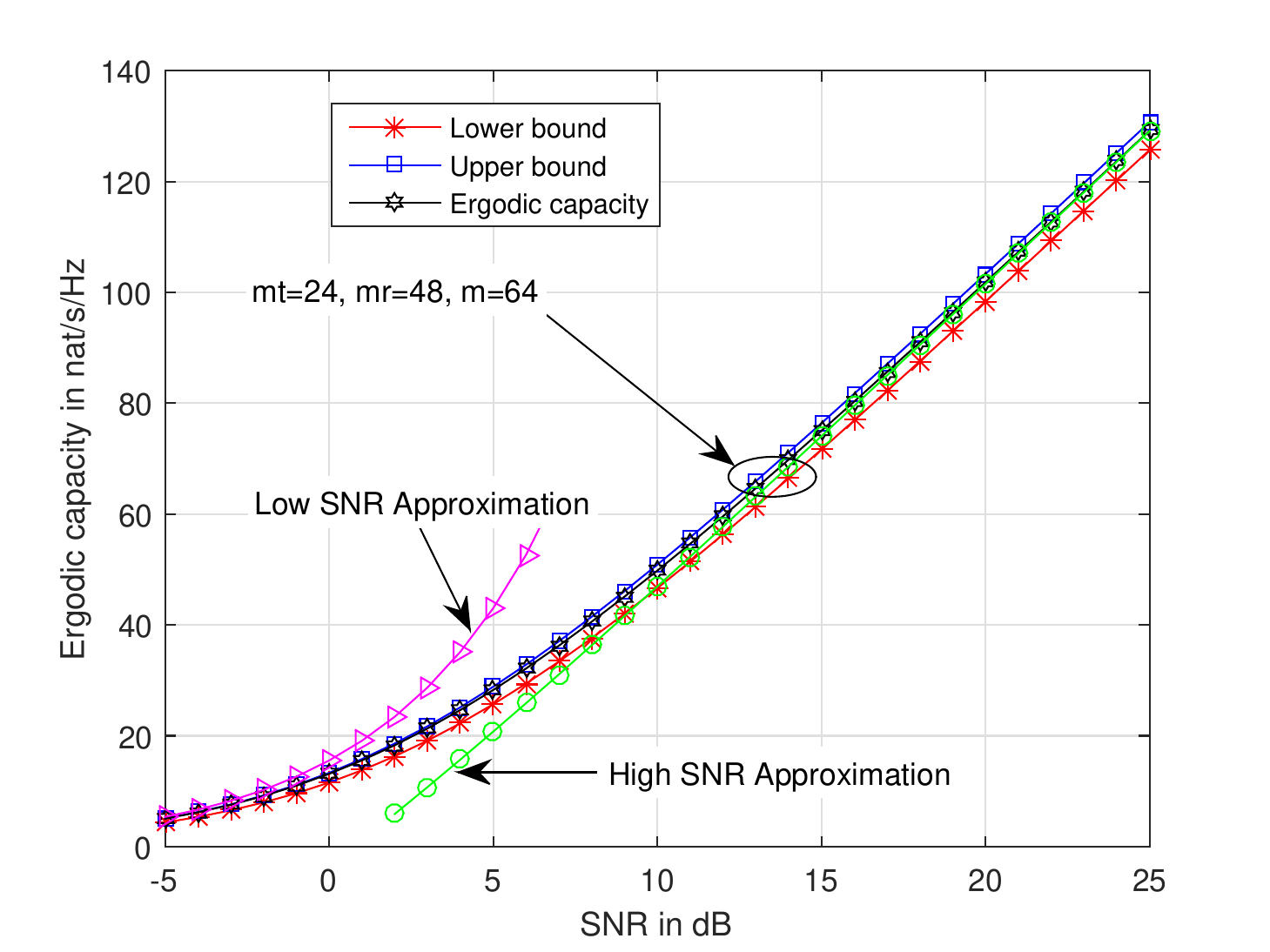}
     %\put(-95,-8){\rotatebox{0}{\mbox{(b)}}}
  \end{minipage}
\end{center}
\caption{{(a)} Comparison of upper bound, lower bound and ergodic capacity in nats per channel use versus SNR in dB when $m_t + m_r$ is larger than the available modes $m$ {(b)} Bounds, upper and lower SNR approximation of the ergodic capacity of the MIMO Jacobi-fading channel where the number of available modes $m=64$ and $m_t+m_r>m$.}
\label{fig:accurate3}
\end{figure}

In this section, we present numerical results to further investigate the resulting analytical equations. The tightness of the derived expressions is clearly visible in Figs.~\ref{fig:accurate1}--\ref{fig:accurate3}.

In Fig.~\ref{fig:accurate1}(a), we have plotted the exact ergodic capacity obtained by computer simulation and the corresponding lower and upper bounds, for the uncorrelated MIMO Jacobi-fading channels, with $(m_t=m_r=2, m=6)$ and $(m_t=4, m_r=10, m=16)$. At very low SNR (typically below 2 dB), the exact curves and the upper bounds are practically indistinguishable. The gaps between the exact curves of the ergodic capacity and the lower bounds considerably vanish in moderate to high SNR (typically above 20 dB). We can observe that the expression in (\ref{Eq3}) matches perfectly with the ergodic capacity expression in (\ref{EC_DFS}). Figure~\ref{fig:accurate1}(b) shows the ergodic capacities of uncorrelated MIMO Jacobi fading channels, and it proves by numerical simulations the validity of the high-SNR regimes lower-bound approximation given in (\ref{Eq4}). Results are shown for different numbers of transmitted/received modes, with $ m = 4$, $m = 8$, and $m=16$. We see that the ergodic capacities approximations are accurate over a large range of high SNR values. Figure~\ref{fig:accurate2}(a) shows the ergodic capacity and the analytical low-SNR upper bound expression in Eq.~(\ref{Eq2}) for several uncorrelated MIMO Jacobi-fading channels configurations. It is clearly seen that our expression is almost exact at very low SNR and that it gets tighter at low SNR as the number of available modes ($m$) increases.

Figure~\ref{fig:accurate2}(b) shows the comparison of the ergodic capacity of the uncorrelated MIMO Jacobi-fading channels and the derived expressions of the upper and lower bounds where the number of available modes is equal to 128. As can be seen in Fig.~\ref{fig:accurate2}(b), the derived upper and lower bounds of the ergodic capacity are close to the exact expression given in (\ref{Eq1}). We verify that our upper and lower bounds give good approximations of the ergodic capacity even for very large number of available modes (\textit{i.e.} $m=128$).

In Fig.~\ref{fig:accurate3}(a), we investigate how close the ergodic capacity is to its upper and lower bounds in cases where $m_t+m_r > m$. We address this particular case using (\ref{EC_DFS_2}). It can be observed that the proposed upper bound on the ergodic capacity is extremely tight for all SNR regimes when $m_r$ is larger than $m_t$. It is important to note that there exists a constant gap between the lower bound and the exact ergodic capacity at all SNR levels. When $m_t$ is larger than $m_r$, such upper and lower bounds are close to ergodic capacity at all SNR regimes. For comparison purposes, we have depicted in Fig.~\ref{fig:accurate3}(b) the ergodic capacity of the MIMO Jacobi-fading channel obtained by computer simulation, the upper/lower bounds and the high/low SNR approximations when the sum of transmit and receive modes, $m_t + m_r$, is larger than the total available modes, $m$. In the high SNR regimes, the ergodic capacity and its high SNR approximation curves are almost indistinguishable. Similarly, we observe that there is almost no difference between the ergodic capacity and its low SNR approximation in the low SNR regions, while there is a significant difference in the high SNR regimes. This difference can be explained by the fact that the first order Taylor's expansion of $\ln\left(1+x\right)$ is not valid for high values of $x$.

\section{Conclusion}
 \label{sec.:conclusion}
In this paper, we derive new analytical expressions of the lower-bound and upper-bound on the ergodic capacity for uncorrelated MIMO Jacobi fading channels assuming that transmitter has no knowledge of the channel state information. Moreover, we derive accurate closed-form analytical approximations of ergodic capacity in the high and low SNR regimes. The simulation results show that the lower-bound and upper-bound expressions are very close to the ergodic capacity.


%\bibliographystyle{IEEEtran}

\end{document}